\documentclass[
  prd,
  twocolumn,
  showpacs,
  preprintnumbers,
  amsmath,amssymb,
  aps,
  nofootinbib,
  superscriptaddress,
]{revtex4-2}

\usepackage{graphicx}
\usepackage{dcolumn}
\usepackage{bm}
\usepackage{soul,xcolor}
\usepackage{hyperref}
\usepackage{orcidlink}
\usepackage{comment}

\newcommand{\fh}{{\em Frequency-Hough }}

\begin{document}

\title{Impact of coalescence signals on the search for continuous gravitational waves with Einstein Telescope}

\author{Elena Codazzo}
\email{elena.codazzo@ca.infn.it}
\affiliation{INFN, Sezione di Cagliari, Cittadella Universitaria, 09042 Monserrato, CA, Italy}
\author{Lorenzo Mirasola\,\orcidlink{0009-0004-0174-1377}}
\email{l.mirasola@uib.cat}
\affiliation{Departament de Física, Universitat de les Illes Balears, IAC3–IEEC, Crta. Valldemossa km 7.5, E-07122 Palma, Spain}
\author{Matteo Di Giovanni\,\orcidlink{0000-0003-4049-8336}}
\email{matteo.digiovanni@sns.it}
\affiliation{Scuola Normale Superiore, Piazza dei Cavalieri, 7, 56126, Pisa, Italy}
\affiliation{INFN, Sezione di Pisa, Largo B. Pontecorvo 3, 56127, Pisa, Italy}
\author{Pia Astone}
\affiliation{INFN, Sezione di Roma, Piazzale Aldo Moro, 2, I-00185 Rome, Italy}
\author{Sabrina D'Antonio}
\affiliation{INFN, Sezione di Roma, Piazzale Aldo Moro, 2, I-00185 Rome, Italy}
\author{Cristiano Palomba}
\affiliation{INFN, Sezione di Roma, Piazzale Aldo Moro, 2, I-00185 Rome, Italy}
\author{Claudia Lazzaro}
\affiliation{Dipartimento di Fisica, Università degli Studi di Cagliari, SP Monserrato-Sestu, km 0.7, I-09042 Monserrato, Italy}
\affiliation{INFN, Sezione di Cagliari, Cittadella Universitaria, 09042 Monserrato, CA, Italy}
\author{Andrea Contu\,\orcidlink{0000-0002-3545-2969}}
\affiliation{INFN, Sezione di Cagliari, Cittadella Universitaria, 09042 Monserrato, CA, Italy}
\author{Alessandro Riggio}
\affiliation{Dipartimento di Fisica, Università degli Studi di Cagliari, SP Monserrato-Sestu, km 0.7, I-09042 Monserrato, Italy}
\affiliation{INFN, Sezione di Cagliari, Cittadella Universitaria, 09042 Monserrato, CA, Italy}
\affiliation{INAF – Istituto di Astrofisica Spaziale e Fisica Cosmica di Palermo, Via U. La Malfa 153, 90146 Palermo, Italy}
\author{Andrea Sanna}
\affiliation{Dipartimento di Fisica, Università degli Studi di Cagliari, SP Monserrato-Sestu, km 0.7, I-09042 Monserrato, Italy}
\affiliation{INFN, Sezione di Cagliari, Cittadella Universitaria, 09042 Monserrato, CA, Italy}
\affiliation{ INAF – Osservatorio Astronomico di Cagliari, Via della Scienza 5, 09047 Selargius, CA, Italy}

\begin{abstract}
The current network of gravitational wave detectors has already revealed hundreds of compact binary coalescences (CBCs), including binary neutron stars, binary black holes, and black hole–neutron star systems. As detector sensitivity improves, the superposition of these signals is expected to form an astrophysical background that becomes increasingly relevant for future observatories. In third-generation detectors, such as the Einstein Telescope (ET), this background will be most prominent at low frequencies, potentially affecting the search for continuous gravitational waves (CWs) from spinning neutron stars.
In this work, we evaluate the impact of the CBC background on CW detection using the \fh pipeline, with a focus on the low-frequency performance in ET sensitivity conditions. Through realistic simulations of the unresolved CBC background, we find that it acts as an additional noise source, most strongly affecting the detection of CW signals around 7\,Hz worsening the FH sensitivity by about 7–10\%.
\end{abstract}


\maketitle

\section{\label{sec:Introduction}Introduction }
Over ten years of observations \cite{BIGONGIARI2026} and four observing runs, the ground-based interferometric detectors Advanced LIGO \citep{aLIGO}, Advanced Virgo \citep{aVirgo}, and KAGRA \citep{somiya2012detector} have detected more than 200 transient gravitational wave (GW) signals from compact binary coalescences (CBCs), including binary black holes (BBHs), binary neutron stars (BNSs), and black hole–neutron star (BHNS) binaries \cite{GWTC1,GWTC2,GWTC21,GWTC3,GWTC4, hawking,pbh25}. These observations have been seminal for a revolution in modern physics that changed the way we observe the Universe \cite{BIGONGIARI2026}: they not only provided definitive proof that compact objects exist in binaries and merge, but also significantly advanced our understanding of compact object populations and properties \cite{abbott2023population,abac2025gwtc}, provided accurate tests of general relativity in the strong field regime \cite{hawking} and explored the mechanisms of binary formation \cite{pbh25}. On top of that, the BNS merger event GW170817 \cite{bns} paved the way for multi-messenger astrophysics \cite{multim}. 

However, several other sources of GW radiation have yet to be detected. Among these are long-lasting, nearly monochromatic continuous waves (CW) emitted by rotating, non-axisymmetric neutron stars (NSs), with GW frequencies related to their spin frequency~\cite{Lasky_2015, WETTE2023102880,riles2023searches}. For these sources the frequency changes much more slowly compared to transient sources, occurring over timescales of years rather than seconds. Since the emission mechanism of CW is closely tied to the internal properties and evolution of NSs, the detection of CWs from these objects would provide unprecedented insights into their yet enigmatic interiors \cite{WETTE2023102880}. It would also offer a chance to study dense matter under conditions different from those in BNS inspirals and mergers, and to perform additional tests of gravitational theory \cite{isi}. Due to the inherently weaker amplitude of CWs compared to the already-detected GW transient sources, searches for CWs from rotating non-axisymmetric NSs are primarily confined to the Milky Way and require extended observation periods to improve the signal-to-noise ratio.

Within the LIGO--Virgo--KAGRA collaboration (LVK), there are several pipelines dedicated to CW searches (see \cite{WETTE2023102880} and references therein) from isolated NSs. These pipelines employ different data analysis methods and have been used in various types of searches - all-sky, directed, narrowband, and targeted - using data from the LVK observing runs, e.g.~\cite{CW1,CW2,CW3,CW4,CW5,abbott2022all, LIGOScientific:2025kei, Steltner_2023, LIGOScientific:2026plm, LIGOScientific:2026qsb}. Through these searches, increasingly stringent upper limits have been set on the maximum ellipticity of neutron stars~\cite{scientific2025search}. Among these methods is the \fh (FH) \cite{astone2014method}, which is the focus of this work. This method employs a particular implementation of the Hough transform (HT) \cite{hough59, hough} to reconstruct the source parameters. 

The successful exploitation of current detectors and the craving for still not detected sources, as in the case of CW from isolated NSs, pushed the scientific community to begin thorough investigations \cite{Maggiore_2020, coba, di2025impact} on the prospects of future, more sensitive GW detectors, like Cosmic Explorer (CE) \cite{CE1, CE2, CE3, DiGiovanni2025_ETCE} and the Einstein Telescope (ET) \cite{et, ET2010, ET2011, ET2020, Iacovelli_2024, DiGiovanni2025_ETCE, abac2025scienceeinsteintelescope}, with the latter being the subject of this work. ET was first proposed in 2010, and the foreseen improvements with respect to current-generation detectors include the extension of the observation bandwidth from the current limit of about 20\,Hz to 2\,Hz and an improvement of the sensitivity by one order of magnitude \citep{ET2020, coba}. The accessible source landscape will be significantly broadened \cite{Maggiore_2020, abac2025scienceeinsteintelescope}

 However, this comes at a cost. Due to the enhanced sensitivity of ET, transient signals in the accessible bandwidth, CBC in particular, will superimpose in time and frequency creating a stochastic astrophysical background \cite{Buonanno2005,RegimbauMandic2008,Regimbau2011,Bellie2024}. This background is expected to dominate at low frequencies and could hinder CW searches by masking or mimicking the detection and reconstruction of weak continuous signals.

In this framework, several works have already investigated the impact of the CBC background on the detectability of the primordial stochastic GW background \cite{zhong2024searching,zhong2025two,sachdev2020subtracting,sharma2020searching}. Nevertheless, a comprehensive study of the influence of this background on CW searches for isolated NSs has not yet been completed. In this work, we conduct for the first time such investigation assuming the ET detector, using the FH pipeline and focusing on the low-frequency band. By simulating GW signals from CBC populations and analyzing their contribution to the overall detector noise, we aim to assess the impact of this background on CW search sensitivity and, possibly, to identify strategies to mitigate its effects in the future.

The paper is organized as follows. 
In Sec.~\ref{sec:CWmorphology}, we discuss the CW emission mechanism and the expected signal at the detector.
Afterwards, Sec.~\ref{sec:Methodology} introduces the tools used for this study, while Sec.~\ref{sec:Noise} describes the data simulation carried out for ET and astrophysical CBC background. 
The obtained results are presented in Sec.~\ref{sec:Results}, with our conclusions drawn in Sec.~\ref{sec:Conclusions} together with some final remarks. Additional details are provided in the Appendix.

\section{\label{sec:CWmorphology}CW signal morphology}

The GW strain produced by a non-axisymmetric neutron star rotating about one of its principal axes can be written in the detector frame as ~\cite{Jaranowski:1999pd}
\begin{align}
\label{eq:cw_signal}
h(t)=& h_0 \sin\xi \,F_+(t;\hat{n},\psi)\,  \frac{1+\cos^2\iota}{2}\, \cos \phi(t) + \nonumber\\
        &h_0 \sin\xi \,F_\times(t;\hat{n},\psi)\, \cos\iota\,\sin \phi(t)
\end{align}
where $\xi$ is the angle between the interferometer arms, $F_{+}$ and $F_{\times}$ are the time-dependent antenna patterns of the detectors, which are a function of the source's sky location $\hat n$ and the polarization angle $\psi;\, h_0$ is the signal amplitude; $\iota$ is the angle between the star total angular momentum vector and the observer's line of sight. Lastly, $\phi(t)$ is the signal phase enclosing the relative motion of source and detector and the intrinsic signal evolution. For a triaxial star rotating about one of its principal moments of inertia, the CW emission is at twice the rotational frequency, namely $f_0 = 2\,f_{\rm rot}$. The signal amplitude, for a source at a distance $d$, can be written as
\begin{equation}
    h_0 =  \frac{4 \pi^2 G}{c^4} \frac{\varepsilon I_{z z} f_0^2}{d}\,,
\end{equation}
with $\varepsilon$ the NS equatorial ellipticity, and $I_{\rm zz}$ the moment of inertia of the star with respect to the principal axis coincident with the rotation axis.

Due to the relative motion between source and the detector, the received frequency $f(t)$ is Doppler shifted and is related to the emitted frequency by~\cite{astone2014method,riles2023searches,WETTE2023102880}
\begin{equation}
    f(t)=f_0(t)\left(1+\frac{\vec{v} \cdot \hat{n}}{c}\right),
\end{equation}
with $\frac{\vec{v} \cdot \hat{n}}{c}$ being the Doppler shift related to the Earth's movement. Here, $\vec{v}=\vec{v}_{\text {rev }}+\vec{v}_{\text {rot }}$, denotes the total velocity of the Earth given by the vectorial sum of revolution ("rev") and rotation ("rot") motions. Additionally, the intrinsic signal frequency $f_0(t)$ varies over time as the NS gradually slows down due to the loss of rotational energy through the emission of electromagnetic and gravitational radiation. Hence, $f_0(t)$ can be expressed as $[9,14]$
\begin{equation}
    f_0(t)=f_0+\dot{f}_0\left(t-t_0\right)\,,
    \label{eq:freq_evol}
\end{equation}
where $\dot{f}_0$ is the first time derivative of the frequency, usually referred to as the spin-down parameter.
Very recently, some all-sky searches~\cite{LIGOScientific:2026plm} have started including a second-order spin-down term in Eq.~\eqref{eq:freq_evol} to account for a more complex dynamic of the signal.

\section{\label{sec:Methodology}Methodology}
To quantify the impact of the CBC background on CW detections, we perform two identical analyses in which we inject a set of artificial CW signals into the data. In the first case, the data consists of simulated ET noise only; in the second, the same CW signals are injected into ET noise that also includes the astrophysical CBC background.
For simplicity, we refer to these two cases throughout the paper as follows:
\begin{itemize}
    \item \textbf{ET0}: analysis using data containing only ET simulated instrumental noise;
    \item \textbf{ETC}: analysis using data containing the same ET noise into which the astrophysical CBC background has been injected.
\end{itemize}
More information on the creation of ET0 and ETC data will be given  in Sec.~\ref{sec:Noise}.

The starting point to create the ET0 and ETC datasets is represented by the so-called Short Fourier Transform Databases (SFDBs)~\cite{astone2005short}.
SFDBs are a collection of Fourier transforms of the data, computed over successive, partially overlapping time segments. In this work, we consider segments of 1024\,s with 50\% overlap.

From the SFDBs, we have computed Band Sampled Data (BSD) files, which contain sub-sampled time series of the data - at 10 Hz sampling rate -, each covering one month and 10\,Hz.\cite{piccinni2018band}. The BSD framework enables flexible, efficient access to subsets of the data. We have generated one month of BSDs covering the [0, 250]\,Hz frequency range. 
These BSDs contain the input data for the Frequency Hough (FH) pipeline~\cite{astone2014method}, where the CW analysis is performed for the two considered cases.

\subsection{\label{sec:FH}The frequency-Hough pipeline }

The FH pipeline~\cite{astone2014method} is a semi-coherent method for detecting unknown CW sources built upon the Hough Transform~\cite{hough, hough59}.
It is a very well-established method, already applied in several CW searches (e.g.~\cite{PhysRevD.100.024004, abbott2022all, LIGOScientific:2026plm, CW1, CW2, CW3, CW4, CW5}).

In this approach, data are divided into segments of length $T_{\textrm{FFT}}$ (known as \textit{coherence time}) that are then processed through Fast Fourier Transforms (FFTs)~\cite{astone2005short}. The segment length is determined by the frequency range under consideration~\cite{Astone:2000jza, piccinni2018band, LIGOScientific:2026plm}. 
In the frequency domain, each chunk of the periodogram is divided by its average spectrum estimation, evaluating the ratio R.
From this equalized spectrum, frequency bins are selected if R is above a given threshold, linked to the per-chunk false-alarm probability, and are local maxima.
After repeating this procedure for all the chunks, the collection of selected peaks forms a time-frequency map usually referred to as \textit{peakmap}~\cite{astone2005short, astone2014method}.

This approach worsens the sensitivity compared to a fully coherent search, where data are analyzed in a single chunk. 
On the other hand, it significantly reduces the computing power required to explore the vast parameter space of all-sky searches~\cite{astone2014method}.
Usually, FH searches explore a certain region in the $f-\dot f$ space, and cover all sky locations from which a CW can be emitted (i.e., ecliptic longitude $\lambda\in [0,2\pi)$, and ecliptic latitude $\beta\in[-\pi/2, \pi/2]$).

The coherence time dictates the number of templates to be explored, as the resolutions in the aforementioned parameters can be expressed as
\begin{align}
    \delta f = \frac{1}{T_{\textrm{FFT}}},&\,\,\, \delta \dot f = \frac{1}{T_{\rm{FFT}} \, T_{\rm obs}}\,,   \notag  \\
    \delta \lambda = \frac{c}{v\,T_{\rm FFT}f \cos\beta},&\,\, \,\delta \beta = \frac{c}{v\,T_{\rm FFT}f \sin\beta}\,,
    \label{eq:resolutions}
\end{align}
where $T_{\rm obs}$ is the total observation time, and $v/c\sim 10^{-4}$ is the Earth's orbital velocity.
Following~\cite{LIGOScientific:2026plm}, we use the new BSD implementation of this pipeline.
This allows for custom $T_{\textrm{FFT}}$ optimized for each frequency band of interest~\cite{piccinni2018band, Astone:2000jza}.

For each point of the sky grid, the peakmap is corrected for the Doppler modulation due to the detector motion.
The resulting peakmap is the input for the FH transform, which maps them onto the source parameter plane $(f,\dot f)$. The output is a two-dimensional histogram where each bin corresponds to a specific combination of intrinsic frequency and spin-down values~\cite{astone2014method}.

From the Hough map, candidates are selected through their critical ratio (CR), which quantifies their statistical significance~\cite{astone2014method, Mirasola:2024kll}:
\begin{equation}
    CR = \frac{x-\mu}{\sigma},
\end{equation}
with $x$ number of counts in a given Hough map pixel, $\mu$ the average across the map, and $\sigma$ its standard deviation.
Usually, for each sky location, the pipeline divides a 1~Hz-band Hough map into 20 intervals and can select up to two candidates from each sub-band, those with the highest CR.
For more information on the candidate's selection, see~\cite{astone2014method}.

To evaluate the proximity between two sets of parameters, a distance metric is defined as:
\begin{equation}\label{eq:distance}\scriptstyle
    d=\sqrt{\left ( \frac{f_1-f_2}{\delta f} \right )^2 + \left ( \frac{\dot f_1-\dot f_2}{\delta \dot f} \right )^2 + \left ( \frac{\lambda_1-\lambda_2}{\delta \lambda} \right )^2 + \left ( \frac{\beta_1-\beta_2}{\delta \beta} \right )^2},
\end{equation}
where $\delta f$,  $\delta \dot f$,  $\delta \lambda$,  $\delta \beta$ are the parameter bin widths  (see Eq.~\eqref{eq:resolutions}), while the labels 1 and 2 refer to the considered pair of candidates.

\section{Data generation\label{sec:Noise}}
In this section, we provide the details of how the data used in our analysis are generated: we simulate the time series for both the expected instrumental noise of ET and the astrophysical background arising from overlapping CBC signals.

The Section is organized as follows: first, in Sec.~\ref{sec:ETnoise}, we briefly describe the pure ET noise generation; then, in Sec.~\ref{sec:CBC}, we outline the procedure to characterize the expected CBC population; in Sec.~\ref{sec:spectral_prop}, we characterize and compare the two obtained time series.

\subsection{\label{sec:ETnoise}Einstein Telescope noise generation}

The ET noise is generated as white Gaussian noise and then colored according to the ET design sensitivity curve used in~\cite{coba}. For this study, we produce a 31-day-long time series of simulated noise, sampled at 512~Hz.

We simulate data for single ET interferometer in its triangular configuration ($\xi=60$\textdegree\, in Eq.~\eqref{eq:cw_signal}) with 10\,km-long arms, complete with both high- and low-frequency detectors.
The detector is located in Sos Enattos, Sardinia, one of the candidate sites~\cite{coba}.

\subsection{\label{sec:CBC}Construction of the astrophysical background}
The populations used to produce the astrophysical background are BBH, BNS, and BHNS systems.
The binary parameters for BBH and BNS are taken from the same catalogs used to produce the results in \cite{coba}. Specifically, BBH parameters are drawn from \cite{mapelli2022cosmic, mapelli2021hierarchical}, which include systems formed through both isolated and dynamical channels, based on Population I and II stellar models.
The parameters for BNS binaries are taken from \cite{santoliquido2021cosmic}, where the populations are derived from stars that evolve in an isolated environment. The same holds for BHNS binaries, whose parameters are drawn from \cite{iorio2023compact}. These CBC population catalogs are also used to produce the results reported in \cite{di2025impact} and \cite{tania2025mock}.

From these catalogs, we obtain datasets containing binaries that merge within one year, with redshift ($z$) distributions extending up to \(z \sim 14\) for all populations. BNS mergers dominate in number, exceeding BHNS events by about two orders of magnitude.
We distribute the merger times of the signals uniformly over the duration of the dataset.

On average, we obtain about 2000 BNS signals per day, compared to roughly 320 BBH and 5 BHNS merger events.

Masses, redshifts, and luminosity distances are taken directly from the catalogs, along with the inclination angle, polarization angle, phase, right ascension, and declination. The spins are set to zero, and the masses are converted to the detector frame by multiplying them by a factor of $(1+z)$. The starting frequency is 2\,Hz and the sampling frequency is 512\,Hz.

BNS signals are long-lived, typically remaining in the ET-sensitive band for more than 10 hours when starting at 2\,Hz, leading  to a stochastic background in which individual contributions cannot be resolved. In contrast,
BBH signals are shorter, lasting at most a few hours. Their larger masses lead to a faster inspiral evolution, so the signal evolves more rapidly in frequency, reducing the time spent within the observational band, and the merging takes place at lower frequencies.
BHNS signals exhibit durations comparable to BNS events but constitute a very small fraction of the total population.

The astrophysical background is generated by simulating individual inspiral waveforms with the TaylorT2 approximant, as implemented in the LIGO Algorithm Library \cite{lalsuite}, and covers the [2, 256] Hz frequency band. Although TaylorT2 does not accurately model the merger and ringdown phases, this is not expected to impact our analysis 
since CW searches are not optimized for short-duration transient signals, which are typically mitigated by cleaning procedures.
We accounted for Earth's rotation by projecting the signals onto the detector noise.

We generate approximately one month of data for all three populations, a choice motivated by the computational cost of both the analysis and the production of realistic noise time series. 
To avoid boundary effects, we trim the edges of the timeseries so that, at any given time, all signals contributing to the data are fully contained within the dataset. This results in 31 days of effective data.

All individual waveforms are then summed to produce a single 31-day time series, which constitutes our astrophysical background from CBC sources. Both this background and the ET noise timeseries are generated with a consistent GPS time reference to ensure temporal coherence.

\subsection{Spectral Properties \label{sec:spectral_prop}}

Each CBC population contributes differently to the total background. Although BBHs represent only about 15\% of the events, their spectral density (ASD) dominates due to their higher strain amplitudes. Their spectra show a steeper slope and terminate at lower frequencies compared to those of BNSs, whose longer inspiral extends to higher frequencies, beyond the range considered here. The astrophysical background we obtained is the sum of the time series from all three populations.

\begin{figure}[!ht]
    \centering
    \includegraphics[width=1\linewidth]{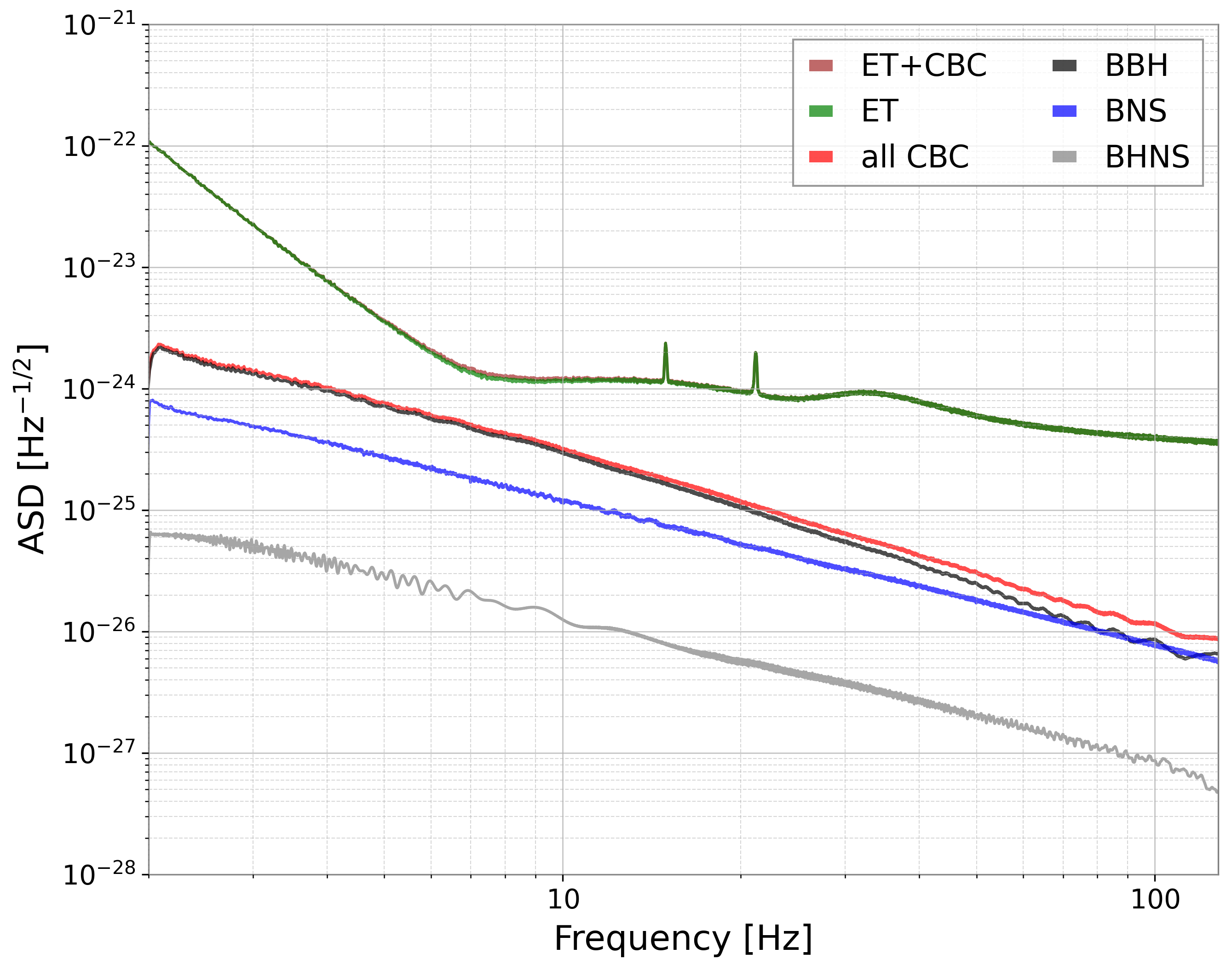}
    \caption{ASDs of the three compact binary coalescence populations in the frequency range [2, 256]\,Hz: BBH (black), BNS (blue), and BHNS (gray). Their sum is shown in red. The green curve corresponds to ET noise alone, while the brown curve shows ET noise plus the CBC background. The two spectra are very similar on this scale, with no significant visible differences, although a small deviation will be discussed in the following. All spectra refer to day 13 of the timeseries and are computed using the Welch method with 128\,s Hann-windowed segments at a sampling frequency of 512\,Hz.}
    \label{fig:ASD}
\end{figure}

Figure~\ref{fig:ASD} shows these ASDs using one day of data: for each population separately and  in red for the combined background. We find that, on this timescale, the spectral estimate is already stable, indicating that one day of data is sufficient to obtain a representative ASD of the stochastic background.

\begin{figure*}[!ht]
    \centering
    \includegraphics[width=0.45\textwidth]{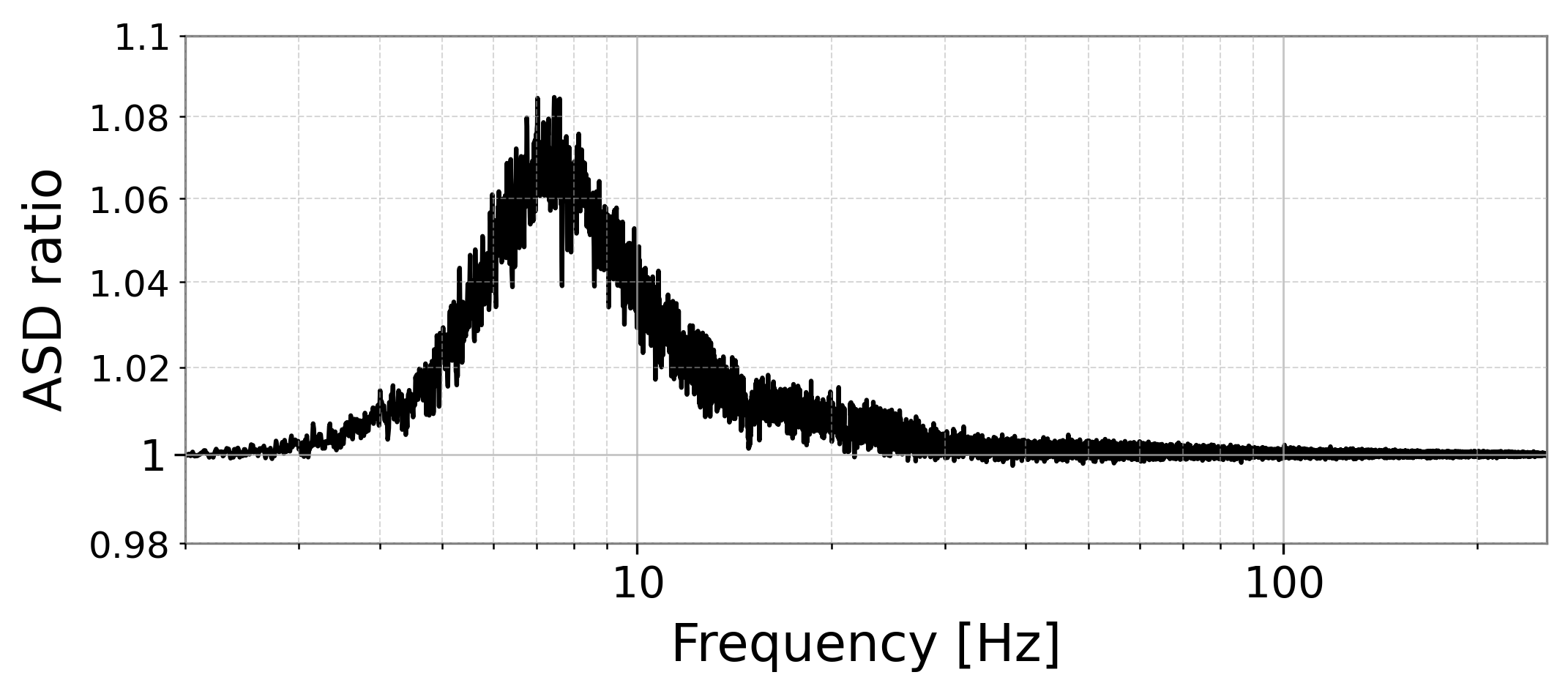}
    \includegraphics[width=0.45\textwidth]{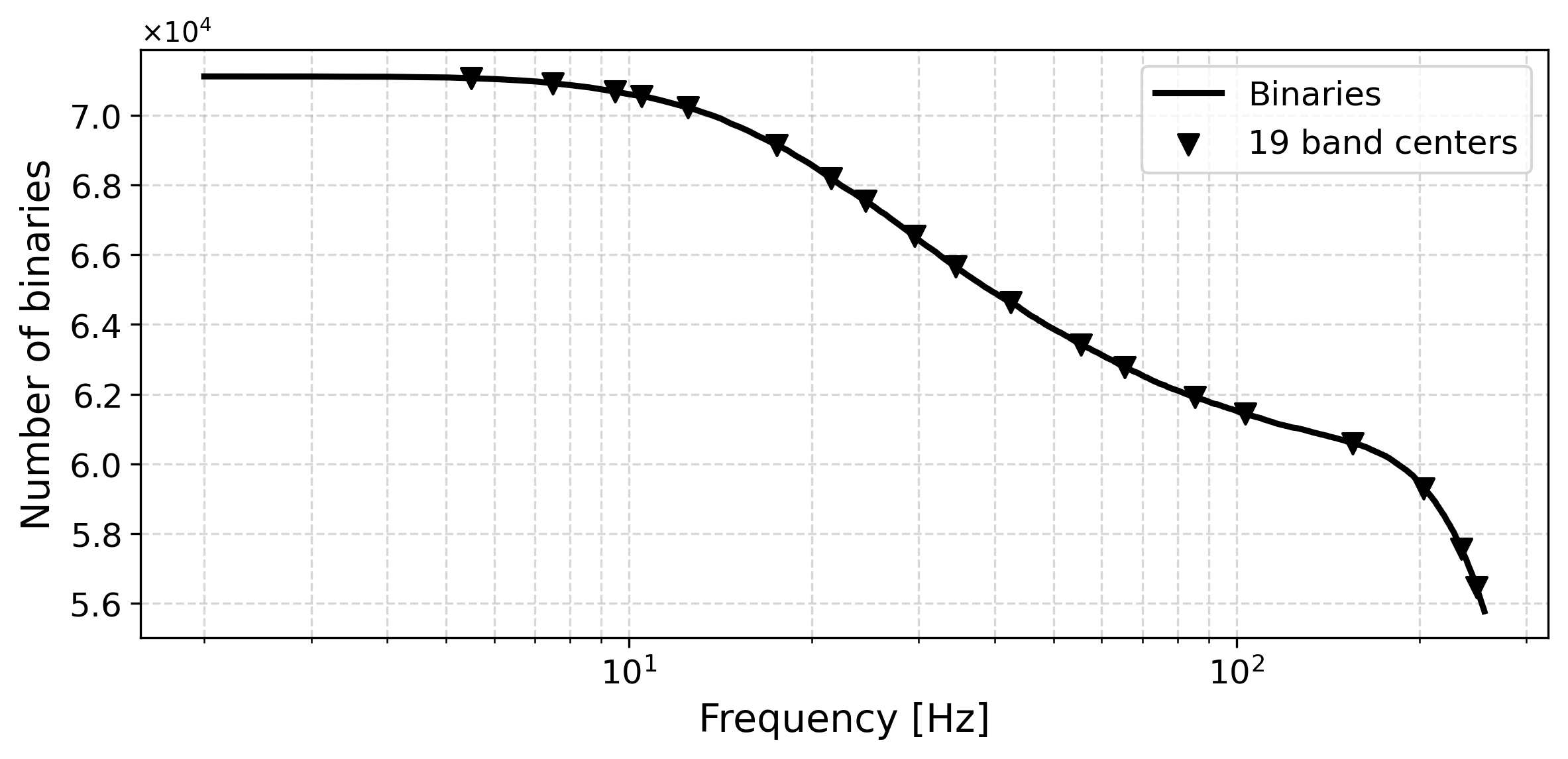}
    \caption{(\textit{left}) Ratio between the ASD of the data containing the CBC background injected into ET simulated noise and the ASD of ET noise alone, for day 13 of the timeseries. The ASDs were estimated using the Welch method with 128\,s Hann-windowed segments at a sampling frequency of 512\,Hz. (\textit{right}) Evolution of the number of inspiral binaries over the simulated month (solid line) as a function of frequency. Inverted triangles indicate the central frequencies of the 19 bands, each 1\,Hz wide, used in the analyses.}
    \label{fig:ASDratio}
\end{figure*}

To assess the impact of this background on ET sensitivity, we injected the full astrophysical background into one month of ET simulated noise. The resulting ASD shows a slight excess over the pure ET noise below 20\,Hz, peaking around 7\,Hz. This effect becomes clearer when computing the ratio between the ASD of ET noise with the CBC background over ET noise alone. This deviation from unity in the ratio spectrum, visible in the left panel of Fig.~\ref{fig:ASDratio}, quantifies the subtle but non-negligible imprint of the astrophysical background on ET data.

In this first study, we do not include additional populations such as population III black holes \cite{santoliquido2023binary} and primordial black holes \cite{ng2022single,luca2020primordial}, which may contribute in particular at high redshift (e.g., z$>$10).  Although their event rates are highly model-dependent and span orders of magnitude \cite{santoliquido2023binary, kinugawa2020chirp,de2020constraints}, they could still contribute to the low-frequency peak we identified.

All waveform generation and signal injections are performed using the \texttt{gwpy} \citep{gwpy, duncan_macleod_2023_7821575} and the \texttt{pycbc} \citep{alex_nitz_2023_7692098} Python packages.

\section{\label{sec:Results}Results }
In this section, we detail the procedure to assess the impact of CBC signals on CW all-sky searches using the FH pipeline~\cite{astone2014method}.
To do so, we generated two datasets as described in Section~\ref{sec:Methodology} and performed two identical analyses using noise ET data alone (ET0), and the other using ET noise combined with the CBC background (ETC).

We selected 19 frequency bands between 5\,Hz and 249\,Hz, each 1\,Hz wide. The starting frequencies of the bands were:
[5, 7, 9, 10, 12, 17, 21, 24, 29, 34, 42, 55, 65, 85, 103, 155, 203, 234, 248]~Hz.
As mentioned in Sec.~\ref{sec:FH}, we tune the $T_{\textrm{FFT}}$ for each frequency band following~\cite{piccinni2018band,LIGOScientific:2026plm}. However, for computational reasons, when the estimated optimal duration exceeded 16384\,s, it was set to $T_{\textrm{FFT}}$=16384\,s. The per-band $T_{\textrm{FFT}}$ are listed in Tab.~\ref{tab:tfft}.

\begin{table}[ht!]
\centering
\begin{tabular}{l r}
Frequency band [Hz] \,\, & \,\, T$_{FFT}$ [s]\\
\hline
{$[5,6]\leq f \leq [35,36]$} & 16384 \\
{[42,43]}          & 15440 \\
{[55,56]}          & 13532 \\
{[65,66]}          & 12464 \\
{[85,86]}          & 10920 \\
{[103,104]}        & 9928  \\
{[155,156]}        & 8108  \\
{[203,204]}        & 7092  \\
{[234,235]}        & 6608  \\
{[248,249]}        & 6420  \\
\hline
\end{tabular}
\caption{Coherence time ($T_{\textrm{FFT}}$) used in each frequency band. For frequencies below 36\,Hz, $T_{\textrm{FFT}}$ was fixed to 16384\,s~\cite{LIGOScientific:2026plm}.}\label{tab:tfft}
\end{table}

In each band, we injected 10 simulated CW signals with equally-spaced frequencies and randomly selected parameters, with a uniform distribution, within the ranges listed in Tab.~\ref{tab:param}.

\begin{table}[!ht]
    \centering
    \begin{tabular}{cc}
    \hline
        $\alpha$ & $[0, 2\pi]$ \\
        $\sin\delta$ & [-1, 1] \\
        $\cos\iota$ & [-1,1]\\
        $\psi$ & [$-\pi/4, \pi/4$]\\
        $\dot{f}$ & $[-10^{-10},\,10^{-10}]$\,Hz/s\\
    \hline
    \end{tabular}
    \caption{Parameters used to generate the CW signals: right ascension ($\alpha$), declination ($\delta$), inclination angle ($\iota$), polarization angle ($\psi$), and spin-down ($\dot{f}$). The values are drawn randomly from uniform distributions within the ranges indicated in the right-hand column. }
    \label{tab:param}
\end{table}

To average the effect of sky localization, we repeated this procedure for 150 different points in the sky.
Additionally, for each frequency band, we use the same set of 10 CW parameters.
To reduce the computational cost, we restrict the analysis to sky points within four bins (see Sec.~\ref{sec:FH}) of the simulated sky location. Indeed, a candidate is considered in coincidence with a simulated signal only if its distance from the injection satisfies $d \le 3$ (see Eq.\ref{eq:distance}); therefore, candidates located farther than four bins in the sky would never be associated with the injected signal. Among the resulting candidates, only those with CR$>$5 are retained, in order to exclude spurious detections likely arising from noise fluctuations \cite{DiGiovanniveto,  CW5, LIGOScientific:2026plm}.

We repeat the procedure for several signal strengths to accurately determine the minimum detectable amplitude at the 95\% confidence level.
We explore amplitudes between [0.7, 1.6] times the theoretical estimation $h_{0,\rm min}$, computed following~\cite{astone2014method, correction_hsens} and accounting for the triangular-shaped detector, for a total of 228000 signal injections distributed over 1200 sky positions.

\begin{figure*}[!ht]
    \centering
    \includegraphics[width=0.45\textwidth]{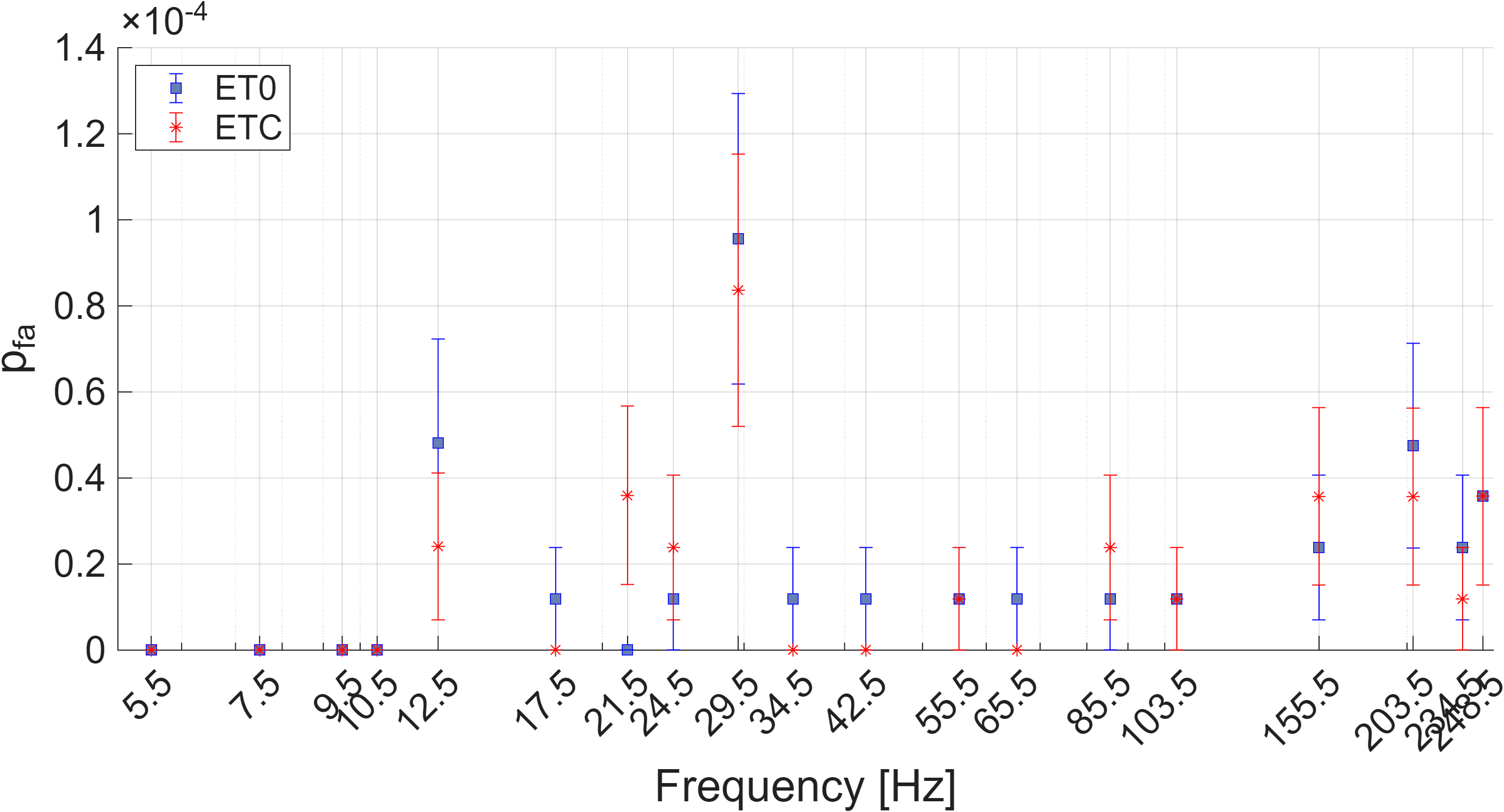}
    \includegraphics[width=0.45\textwidth]{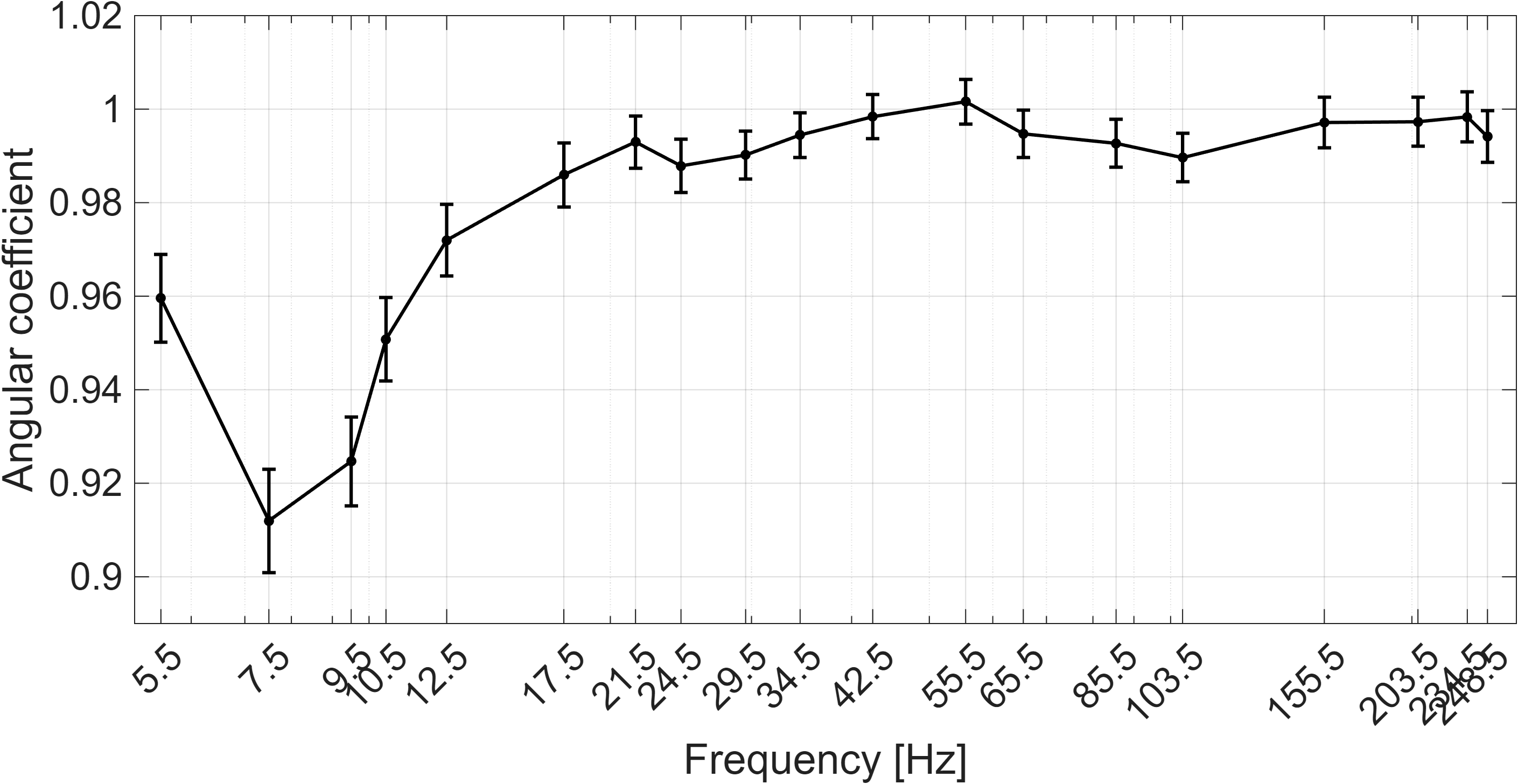}
    \caption{(\textit{left}) False alarm probability ($p_{\rm fa}$) as a function of frequency, for ET0 (dark blue squares) and ETC (dark red stars). The $p_{\rm fa}$ is estimated by running the analyses without any CW injections, such that every recovered candidate corresponds to a false alarm. In both cases, the $p_{\rm fa}$ remains smaller than 0.02\%. (\textit{right}) Comparison of CR from ET0 and ETC analyses for a studied injected amplitude. We perform a linear fit and report its slope as a function of frequency. Error bars indicate the 95\% confidence intervals. Values smaller than unity show a reduction in the recovered CR; thus, values below one indicate a detected effect in the presence of CBCs.}
    \label{fig:FAP}
\end{figure*}

The background noises used in the ET0 and ETC analyses were generated as described in Section \ref{sec:Noise}. There, we showed that the contribution of CBCs to the ET sensitivity curve peaks around 7–8\,Hz. This behavior mainly arises because, at low frequencies, a higher overlap of signals occurs, whereas at higher frequencies only lighter systems contribute. The right panel of Fig.~\ref{fig:ASDratio} shows the distribution of inspiral binaries as a function of frequency over the considered month, with the central frequencies of the 1\,Hz wide bands also indicated by triangle markers. In the following, we show that this frequency-dependent behavior of the CBC background produces measurable differences between the ET0 and ETC analyses, affecting the detection efficiency of CW signals.

\subsection{False-alarm probability\label{sec:FAP}}
To estimate the false-alarm probability ($p_{\rm fa}$), we repeat the same procedure described before, with both ET0 and ETC data, but without injecting any signal (i.e. injection with null amplitude).
In this case, any recovered candidate, in coincidence with the set of ``injection parameters'' and CR$>$5 is, by construction, a false positive. This threshold is consistent with the standard procedure in all-sky searches with real data, which follows up only candidates above the threshold due to computational limitations~\cite{abbott2022all, DiGiovanniveto}. The $p_{\rm fa}$ is computed as
\[
\mathrm{p_{\rm fa}} = \frac{N_{\mathrm{false}}}{N_{\mathrm{points}}}, 
\]
where $N_{\mathrm{false}}$ is the number of recovered candidates and $N_{\mathrm{points}}$ is the total number of grid points explored in each frequency band during the search.

The results are shown in the left panel of Fig.~\ref{fig:FAP}. We find that the $p_{\rm fa}$ remains very small, below 0.02\% across all frequencies, for both the ET0 and ETC cases, confirming the robustness of the pipeline.

\subsection{CW detection\label{sec:COICI}}

Following the procedure described above, we evaluate the CW sensitivities of the FH pipeline (see Sec.~\ref{sec:FH}) by injecting simulated CW signals into both the ET0 and ETC dataset. As suggested by the behavior of the CBC contribution in the ASD ratio of Fig.~\ref{fig:ASDratio}, we expect its impact to be most relevant around $f \sim 7$~Hz. 

The FH pipeline~\cite{astone2014method} selected sets of candidates in coincidence with the injected signals according to Eq.~\eqref{eq:distance} (see Sec.~\ref{sec:FH}). Only candidates with CR$>$5 are retained, following the standard selection used in all-sky searches~\cite{DiGiovanniveto, CW5, LIGOScientific:2026plm}.
We compare candidates obtained in the ET0 and ETC analyses by pairing those with the highest CR associated with the same injection.
The comparison shows a systematic reduction of CR values in the presence of the CBC background, with stronger effects at low frequencies.
This is readily shown in the right panel of Fig.~\ref{fig:FAP}, where we report the angular coefficient of a linear fit in the CR$_{\rm ET0}$ \textit{vs} CR$_{\rm ETC}$ plane.
Angular coefficients lower than one indicate smaller CRs in the ETC analysis.

Specifically, Figure~\ref{fig:FAP} shows the results for the $h_0 = h_{0\,,\rm min}$ amplitude case, presenting the slope of the linear fits as a function of frequency, with error bars at the 95\% confidence intervals.
A clear trend emerges: at lower frequencies, the CBC background appears to reduce the CR values in the ETC analysis with respect to the ET0 analysis, with the most significant deviation from unity occurring in the [7,8]\,Hz band. This is in agreement with the peak shown in Fig.~\ref{fig:ASDratio}. At higher frequency bands, the slope approaches unity, indicating a negligible impact of the CBC background. A representative fit for one frequency band is shown Appendix~\ref{app:CR_fit}.

As a further validation of the behavior observed in the right panel of Fig.~\ref{fig:FAP}, we empirically estimate the amplitude at which 95\% of the injected signals are recovered, $h_0^{95\%}$.
For each studied amplitude and frequency, we count the number of detected injections, i.e., those having at least one candidate within $d\leq3$ and CR$>$5.
The $h_0^{95\%}$ values are then interpolated using a sigmoid fit, as described in App.~B7 in~\cite{LIGOScientific:2026plm} through a sigmoid fit.

\begin{figure}[!ht]
    \centering
    \includegraphics[width=0.5\textwidth]{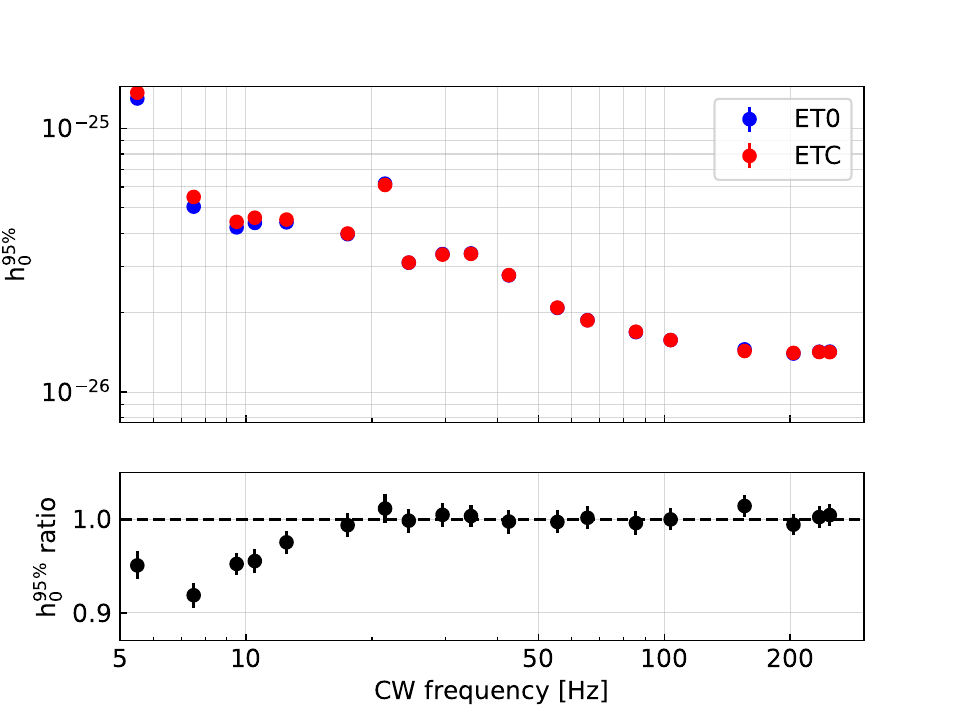}
    \caption{(\textit{Upper panel}) Empirical upper limits at the 95\% confidence level as a function of frequency, see text for more details on their calculation. (\textit{Lower panel}) Ratio of ET0 to ETC upper limits. A clear discrepancy is observed in the low-frequency region, where the CBC background is expected to affect CW searches more strongly. At higher frequencies, the upper limits cluster around unity, as expected.
    }
    \label{fig:hul}
\end{figure}

The results are shown in Fig.~\ref{fig:hul} and are consistent with those reported in the right panel in Fig.~\ref{fig:FAP}. In the low-frequency region, ET0 $h_0^{95\%}$ are better (i.e., smaller) than ETC ones: in this regime, due to the presence of CBCs, the upper limits are worsened, at most, by $\sim7-10\%$ compared to the ET0 scenario, in agreement with the largest ASD variation reported in Fig.~\ref{fig:ASD}.
At higher frequencies ($f\geq 17$~Hz), the two analyses converge, with consistent $h_0^{95\%}$ values. Their ratio, shown in the lower panel of Fig.~\ref{fig:hul}, is also consistent with unity within uncertainties.

In this study, we have neglected potential mitigation strategies for reducing the effect of the CBC background.
For for stochastic GW background searches, examples are given in~\cite{zhong2024searching, zhong2025two, sachdev2020subtracting, sharma2020searching}, where loud CBCs are subtracted in the time domain before proceeding with the analysis.
A detailed study is needed to evaluate the effectiveness of this procedure in the context of CW analyses.
As a result, our findings represent a worst-case scenario for the FH all-sky pipeline, where the impact of the additional astrophysical foreground remains at the level of a few to 10\% in terms of upper limits. Such a variation, while moderate, may still have a significant effect on the required observation time and on the detectability of CW signals, whose intrinsic amplitudes are not known a priori.

\section{\label{sec:Conclusions}Conclusions }

In this work, we have investigated the influence of overlapping CBC signals on CW all-sky searches, using the FH pipeline as a test bed, in the ET era.
To this purpose, we constructed a realistic astrophysical background from synthetic populations of BBH, BNS, and BHNS systems \cite{mapelli2022cosmic, mapelli2021hierarchical,santoliquido2021cosmic,iorio2023compact}. We have shown that frequencies below 20\,Hz are mostly affected by this foreground, which introduces an excess in the noise level with a peak contribution around 7\,Hz.
We then performed an injection campaign consisting of two analyses, one considering ET noise alone and the other including the astrophysical CBC background, finding a consistent trend in the FH results.
We demonstrated how, if not mitigated, this additional source of noise can affect CW searches. Specifically, the FH sensitivity is worsened by 7–10\% in the case in which CBCs are included in the simulation.

Although this analysis has been performed using a single detector from the triangular configuration of ET, we do not expect the main conclusions to change significantly when considering the alternative 2L configuration. Previous studies (\citep{coba}) have shown that different detector configurations lead to comparable sensitivities for semi-coherent all-sky searches with the FH pipeline, with differences mainly related to the arm length. Therefore, while quantitative sensitivity variations may arise, the qualitative impact of the unresolved CBC background should remain unchanged. Nevertheless, further tests in different configuration are need in the future to confirm this expectation.

In light of these results, mitigation strategies will be crucial for future CW searches applied to new-generation detectors to avoid being limited by this astrophysical source.
Methods developed in the context of stochastic background searches, such as the subtraction of loud CBC~\cite{zhong2024searching, zhong2025two, sachdev2020subtracting, sharma2020searching}, offer valuable insights. Developing dedicated techniques to suppress the effect of overlapping CBC events in CW analyses will be an important direction for future work.


\section*{Acknowledgments}
E.C. acknowledges support from the ET Start-up project, funded by the University of Cagliari under the ‘Start Up’ research call (D.M. 737/2021), within the framework of interdisciplinary research initiatives supported by the Italian National Recovery and Resilience Plan (PNRR)
and Regione Autonoma della Sardegna (CUP I73C24000610002).

All the authors acknowledge the Istituto Nazionale di Fisica Nucleare (INFN) for the support.

L.M. acknowledges the support by the Universitat de les Illes Balears (UIB)  with funds from the Programa de Foment de la Recerca i la Innovació de la UIB 2024-2026 (supported by the yearly plan of the Tourist Stay Tax ITS2023-086); the Spanish Agencia Estatal de Investigación grants PID2022-138626NB-I00, RED2024-153978-E, RED2024-153735-E, funded by MICIU/AEI/10.13039/501100011033 and the ERDF/EU; and the Comunitat Autònoma de les Illes Balears through the Conselleria d'Educació i Universitats with funds from the ERDF (SINCO2022/18146-Plataforma HiTech-IAC3-BIO) and by COST action SCALES CA24139, supported by COST (European Cooperation in Science and Technology).

This material is based upon work supported by NSF's LIGO Laboratory, which is a major facility fully funded by the National Science Foundation.

E.C. is grateful to F. Santoliquido and F. Iacovelli for providing useful information on the population catalogs used in this study.

\appendix

\section{CR fit\label{app:CR_fit}}

\begin{figure}
    \centering
    \includegraphics[width=0.8\linewidth]{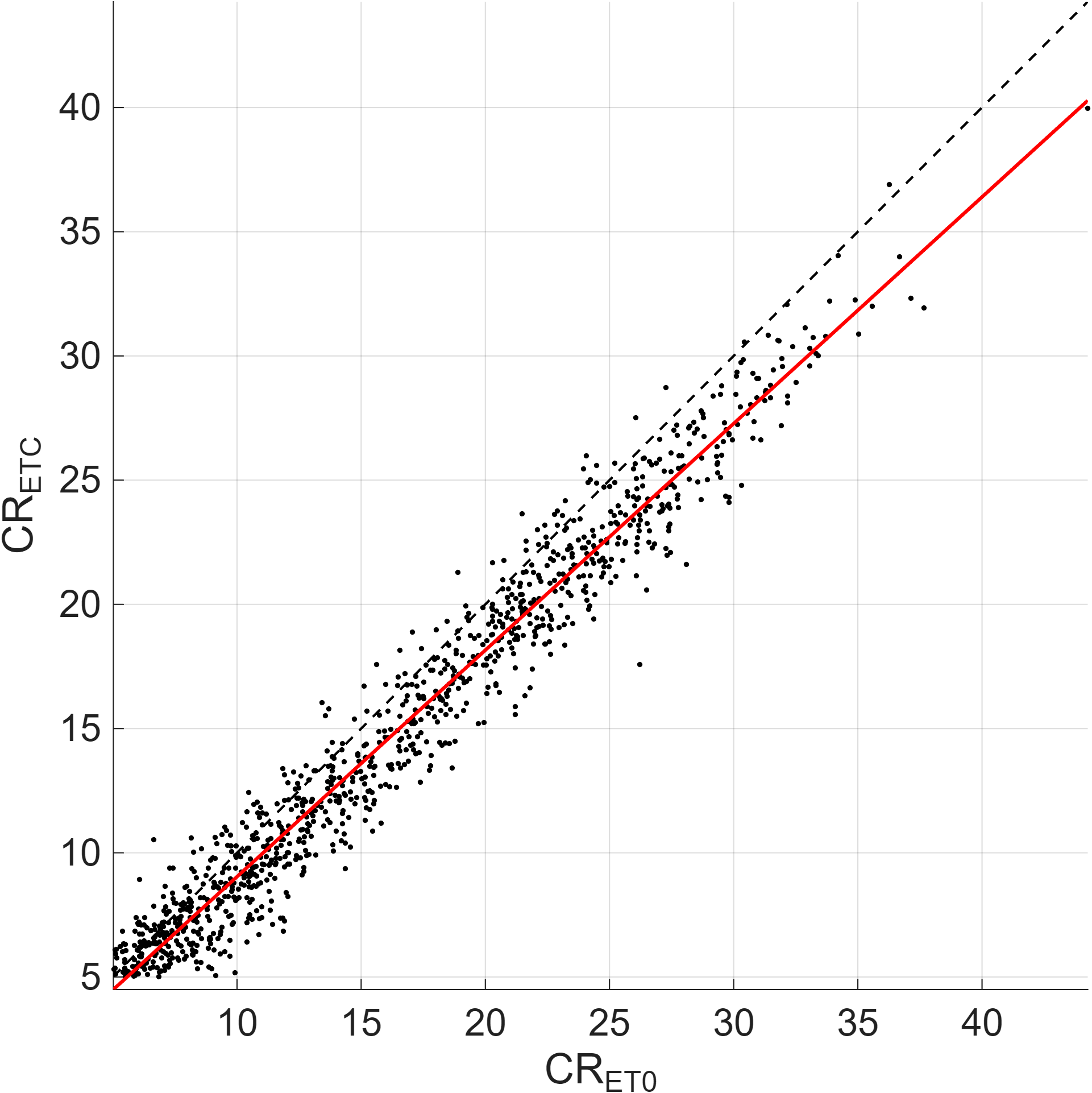}
    \caption{Example of the linear fit between CR$_{ET0}$ and CR$_{ETC}$ for the [7,8]\,Hz frequency band. The red line represents the best-fit linear relation, and the dashed line shows the bisector. The fitted slope $0.912\pm0.006$ indicates a reduction of CR in the ETC analysis due to the presence of the CBC background.}
    \label{fig:fit_7Hz}
\end{figure}

As an illustrative example, Figure \ref{fig:fit_7Hz} shows the linear fit between the CR values obtained in the ET0 and ETC analyses for the [7,8]\,Hz frequency band. Each point represents a pair of candidates associated with the same injected CW signal. The red line corresponds to the best-fit linear relation, while the dashed line indicates the identity line.
The fitted slope is $0.912\pm0.006$, confirming that the CR values in the ETC case are systematically lower than those in ET0, consistent with the larger CBC background impact observed at this frequency.

\nocite{*}
\bibliography{aipsamp}

\end{document}